\documentclass[aps,prx,twocolumn,superscriptaddress, floatfix, showpacs, showkeys]{revtex4-1}

\usepackage{graphicx}
\usepackage{epsfig}
\usepackage{charter}
\usepackage{amssymb}
\usepackage{setspace}
\usepackage{amsmath}
\usepackage{csquotes}
\usepackage{float}
\usepackage {braket}
\usepackage{epstopdf}
\usepackage{subfigure}
\usepackage{amsmath}
\usepackage{color}
\usepackage[ugly]{units}

\usepackage[bookmarks=false]{hyperref}

\usepackage{hyperref}
\usepackage{xcolor}
\usepackage{xfrac}
\definecolor{dark-red}{rgb}{0.4,0.15,0.15}
\definecolor{dark-blue}{rgb}{0.15,0.15,0.4}
\definecolor{medium-blue}{rgb}{0,0,0.5}
\hypersetup{
    colorlinks, linkcolor={dark-red},
    citecolor={dark-blue}, urlcolor={medium-blue}
}

\begin{document}

\title{Robustness of superconductivity to structural disorder in Sr$_{0.3}$(NH$_2$)$_{y}$(NH$_3$)$_{1-y}$Fe$_2$Se$_2$}

\author{F. R. Foronda}
\email{francesca.foronda@physics.ox.ac.uk}
\affiliation{Oxford University Department of Physics, Clarendon Laboratory, Parks Road, Oxford, OX1 3PU, United Kingdom}

\author{S. Ghannadzadeh}
\affiliation{Oxford University Department of Physics, Clarendon Laboratory, Parks Road, Oxford, OX1 3PU, United Kingdom}
\affiliation{High Field Magnet Laboratory, Institute for Molecules and Materials, Radboud University, 6525 ED Nijmegen, The Netherlands}

\author {S. J. Sedlmaier}
\affiliation{Department of Chemistry, University of Oxford, Inorganic Chemistry Laboratory, South Parks Road, Oxford, OX1 3QR, United Kingdom}

\author{J. D. Wright}
\affiliation{Oxford University Department of Physics, Clarendon Laboratory, Parks Road, Oxford, OX1 3PU, United Kingdom}

\author {K. Burns}
\affiliation{Department of Chemistry, University of Oxford, Inorganic Chemistry Laboratory, South Parks Road, Oxford, OX1 3QR, United Kingdom}

\author{S. J. Cassidy}
\affiliation{Department of Chemistry, University of Oxford, Inorganic Chemistry Laboratory, South Parks Road, Oxford, OX1 3QR, United Kingdom}
\affiliation{Diamond Light Source Ltd., Harwell Science and Innovation Campus, Didcot OX11 0DE,
United Kingdom}

\author{P. A. Goddard}
\affiliation{Department of Physics, University of Warwick, Gibbet Hill Road, Coventry CV4 7AL, United Kingdom}

\author{T. Lancaster}
\affiliation{Durham University, Centre for Materials Physics, South Road, Durham, DH1 3LE, United Kingdom}

\author{S. J. Clarke}
\affiliation{Department of Chemistry, University of Oxford, Inorganic Chemistry Laboratory, South Parks Road, Oxford, OX1 3QR, United Kingdom}

\author{S. J. Blundell}
\email{s.blundell@physics.ox.ac.uk}
\affiliation{Oxford University Department of Physics, Clarendon Laboratory, Parks Road, Oxford, OX1 3PU, United Kingdom}

\date{\today}

\begin{abstract}
The superconducting properties of a recently discovered high $T_{\rm c}$ superconductor, Sr/ammonia-intercalated FeSe, have been measured using pulsed magnetic fields down to $\unit[4.2]{K} $ and muon spin spectroscopy down to 1.5 K. This compound exhibits intrinsic disorder resulting from random stacking of the FeSe layers along the $c$-axis that is not present in other intercalates of the same family. This arises because the coordination requirements of the intercalated Sr and ammonia moieties imply that the interlayer stacking (along $c$) involves a translation of either ${\bf a}/2$ or ${\bf b}/2$ that locally breaks tetragonal symmetry. The result of this stacking arrangement is that the Fe ions in this compound describe a body-centred tetragonal lattice in contrast to the primitive arrangement of Fe ions described in all other Fe-based superconductors. In pulsed magnetic fields the upper critical field $H_{\text{c2}}$ was found to increase upon cooling with an upwards curvature that is commonly seen in type-II superconductors of a multi-band nature. Fitting the data to a two-band model and extrapolation to absolute zero gave a maximum upper critical field ${\mu_0H_{\rm c2}(0)}$ of ${33(2)\,\rm T}$. A clear superconducting transition with a diamagnetic shift was also observed in transverse-field muon measurements at ${T_{\text c}\approx36.3(2)\,\rm K}$. These results demonstrate that robust superconductivity in these intercalated FeSe systems does not rely on perfect structural coherence along the $c$-axis.
\end{abstract}

\maketitle

\section{Introduction}
The discovery of superconductivity in iron-based materials \cite{KamiharaJACS130} has led to a new family of systems with substantial structural variations, but all compounds are composed of Fe$B$ layers (where ${B= \rm As, Se, Te, P}$ or some mixture) and the variations arise from the way these layers are assembled and which atoms are included between them \cite{PaglioneNatPhys6, HirschfeldRPP74, JohnstonAdPhys59, DagottoRevModPhys85}. Increasing the inter-layer separation in FeSe has been found to give rise to a dramatic effect on the superconducting transition temperature $T_{\rm c}$. For example, pure FeSe (${T_{\rm c}=8.5\,\rm K}$ \cite{McQueenPRB79}) can be intercalated with alkali metal ions and ammonia \cite{MaziopaJPCM24, BurrardNatMat12, YingSciRep2} or other organic molecules \cite{NojiPhysicaC504, MaziopaJPCM24} to produce new superconductors with transition temperatures of up to around 45\,K. Similar high $T_{\rm c}$ behavior can be induced using metal hydroxides as the spacer layer \cite{SunInorgChem54, PachmayrAngChem54, LuNatMat14, HatakedaJPSJ82}. This trend of increase in $T_{\rm c}$ with increasing layer separation does not continue indefinitely \cite{NojiPhysicaC504}, and this effect can be rationalized through first-principle calculations \cite{GuterdingPRB91}. However, the effect of controlled structural disorder on $T_{\rm c}$ has not been so closely examined. We have identified an intercalated FeSe compound in which random stacking of well-defined layers results in a paracrystalline structure. In this paper we demonstrate that the superconducting state is nevertheless robust. 

\begin{figure*}[tp]
\centering
\includegraphics*[width=0.9\linewidth]{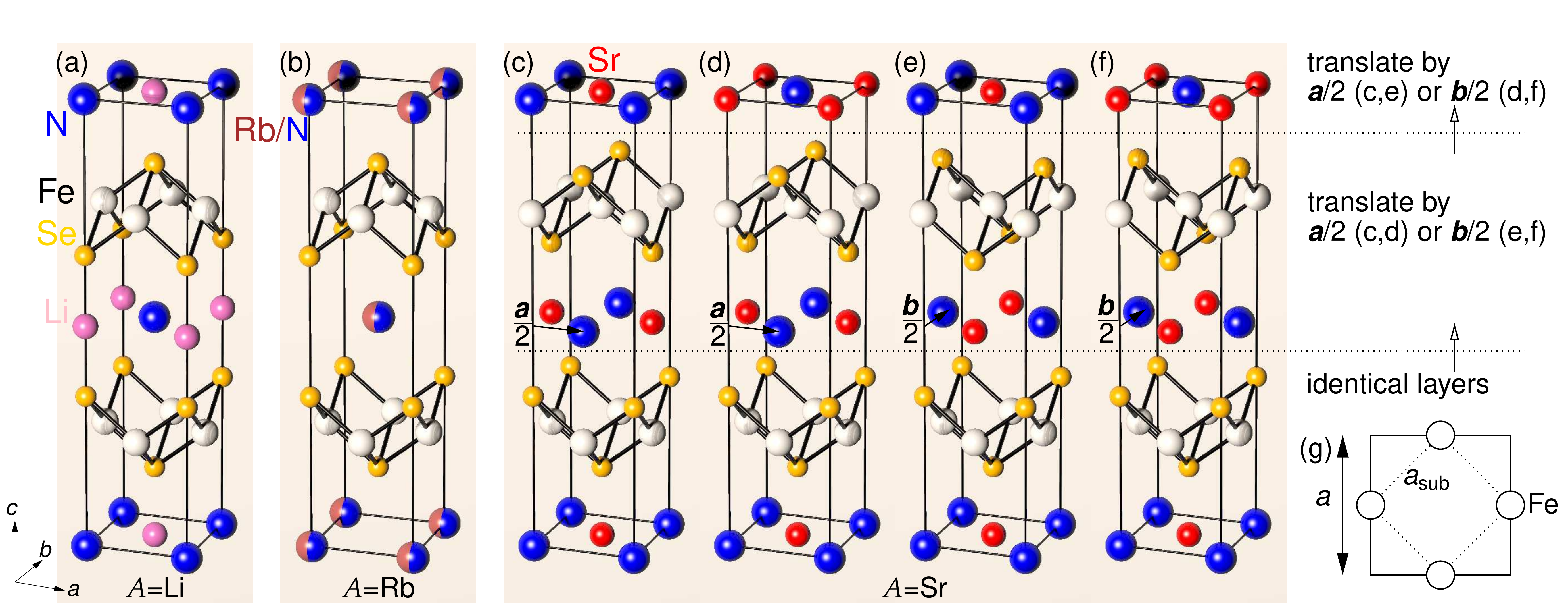}
\caption{Crystal structure of the FeSe intercalates family in which the spacer layer consist of ammonia with (a) Li,  (b) Rb, or (c)--(f) Sr atoms (hydrogen omitted for clarity). In the Li and Rb intercalates the Fe atoms (gray spheres) in adjacent layers occupy a primitive  $a_{\rm sub}\times a_{\rm sub}\times c_{\rm sub}$ sublattice, where ${c_{\rm sub}=c/2}$ is the distance between adjacent Fe layers (${\sim8.3\,\rm\AA}$ for the Li intercalate). In the Rb case, the metal ions and ammonia molecules share the same site with 50\% occupancy (split pink/blue spheres). In the Sr case, the Fe sublattice is a body-centered tetragonal unit cell that is elongated by a factor of two in the $c$ direction and the structure contains random stacking faults. The stacking faults are illustrated in Figures (c)--(f) as follows. The dotted lines represent where the stacking faults occur, with the section below the lower dotted line being identical in all four diagrams.  Above the lower dotted line, the Fe-Se units are translated relative to those in the lower layer by half a unit cell along either the $a$ or $b$ direction. At the upper dotted line, the same translation along $a$ or $b$ occurs again, so that with $n$ dotted lines there would be $2^n$ possible configurations.  With the two dotted lines shown, Figures (c)--(f) demonstrate the four possible stacking combinations when starting from an identical base layer. The position of the Fe sublattice relative to the unit cell in a single layer, as viewed along the c-axis is shown in (g).}
\label{SrFeSe_unitcell}
\end{figure*}

The compound Sr$_{x}$(NH$_2$)$_{y}$(NH$_3$)$_{1-y}$Fe$_2$Se$_2$ (${x=0.3}$, ${0.2\leq\,y\,\leq\,0.6}$) belongs to a family of layered intercalates $A_{\rm x}$(NH$_{2}$)$_{y}$(NH$_{2}$)$_{1-y}$Fe$_2$Se$_2$ ($A= $\, Li, Na, K, Rb, Cs, Ca, Sr, Ba, Eu and Yb). In these materials bulk superconductivity occurs in anti-PbO-type FeSe layers composed of edge-sharing FeSe$_4$ tetrahedra which are separated by metal ions, amide ions and ammonia molecules \cite{BurrardNatMat12, ScheidtTEPJB85, YingSciRep2, SedlmaierJACS136}. Fig.\ref{SrFeSe_unitcell} shows the various structures and the conventional unit cell with ${a=b\approx3.8\,\rm\AA}$ and ${c=16.5\text{--}17.4\,\rm\AA}$. These compounds remain tetragonal down to low temperatures and do not exhibit an orthorhombic distortion. Neutron scattering and X-ray diffraction measurements have revealed some structural differences related to the size of the cation used; when it is small ($A=\rm Li$), Fe ions in adjacent layers occupy the same primitive tetragonal sublattice \cite{BurrardNatMat12}, as shown in Fig.\ref{SrFeSe_unitcell}a. For larger cations (${A=\rm K, Rb}$), the arrangement of Fe and Se ions is unchanged, but the larger electropositive cations share the same sites as the amide and ammonia moieties in the body-centred position of the primitive Fe sublattice (see Fig.\,\ref{SrFeSe_unitcell}b). As we describe in detail in section III the Sr case presents a scenario in which the coordination requirements of the Sr cation and the amide or ammonia moieties are best satisfied by an arrangement of adjacent FeSe layers that results in a random stacking of these layers along the $c$-axis, which may be described as paracrystalline. We will demonstrate through SQUID magnetometry, pulsed magnetic fields and muon spin rotation ($\mu$SR) measurements that, despite this unusual structural disorder, superconductivity remains robust with a $T_{\rm c}\approx\!36\,\rm K$ that is significantly higher than that of a parent compound.

\section{Synthesis}
In an argon filled glovebox, finely ground FeSe powder (1.970 g, 14.611 mmol) synthesized from the elements as described elsewhere \cite{BurrardNatMat12} and pieces of strontium metal (0.640 g, 7.304 mmol, ALFA) were placed in a thick-walled glass Schlenk tube capable of withstanding an internal pressure of over 15 bar. The tube was evacuated and cooled to $-78$ \textdegree C with an isopropanol/dry ice cooling bath. While cold, around 50 ml of liquid ammonia was condensed into the Schlenk tube and the Sr metal was dissolved in the liquid ammonia to produce a blue solution. The valve on the Schlenk tube was closed, isolating it from the line and the reaction mixture was allowed to warm to room temperature and stirred for $24\, \rm h$. After the reaction, it was cooled to $-78$ \textdegree C again, allowing the valve to the line to be opened and the ammonia to evaporate off via a mercury-filled bubbler while letting the Schlenk tube slowly warm to room temperature. At the end of this process the Schlenk tube was placed under dynamic vacuum for 2 min and brought into the glovebox. The product was isolated as a very fine black powder. Samples were prepared using both normal and deuterated ammonia. For further measurements, the samples were handled under an inert gas atmosphere at all times.

\begin{figure*}[t]
\centering
\includegraphics*[width=1\linewidth]{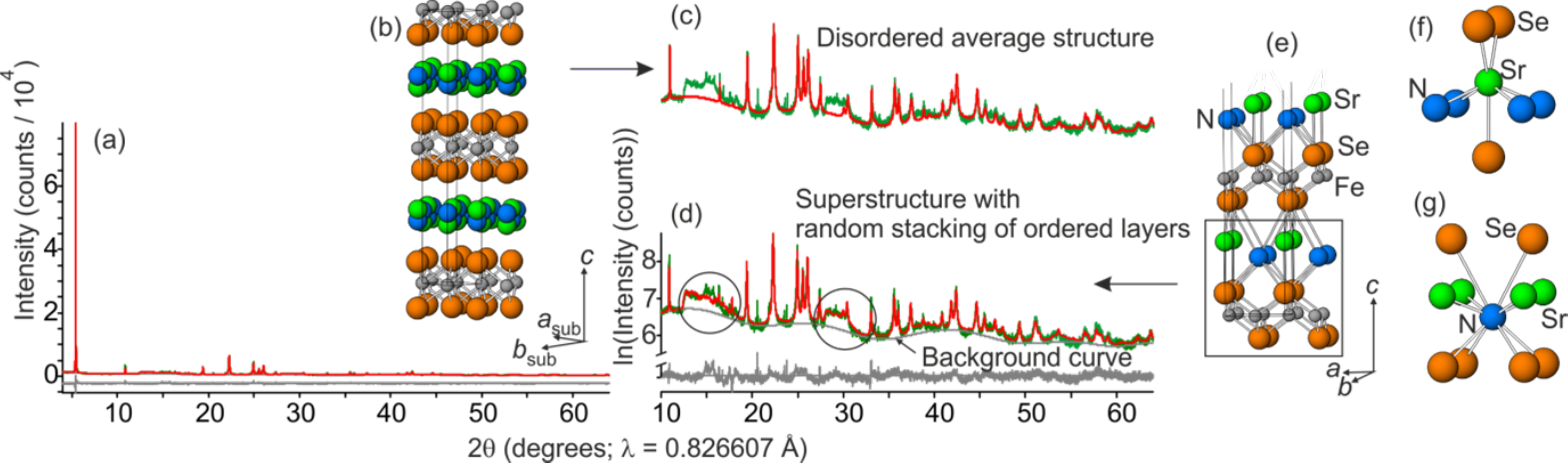}
\caption{Refinements against I11 powder diffraction data showing experimental (green) and calculated (red line) patterns as well as the difference (gray line). The full pattern is shown in (a) with magnifications of the diffractogram showing calculations from two different models in (c) and (d) (magnifications are shown with the same horizontal scale as (a) but with a logarithmic intensity scale to emphasize the weak features in the data). The full pattern is dominated by sharp Bragg reflections which can be accounted for by the disordered average structure shown in (b) with space group {\it I4/mmm} with unit cell ${a_{\rm sub}\times a_{\rm sub}\times c}$. This model assumes full occupancy of the Fe sites that describe a body centered tetragonal unit lattice, but 50\% occupancy of the selenide sites and also partial occupancy of the N and Sr sites. However it fails to account for regions of diffuse scattering, as shown in (c). (d) shows an improved fit to the diffuse scattering (ringed) when using a model in which chemically realistic layers (boxed region in diagram (e)) with basal lattice parameters $a=b=\sqrt{2}\times a_{\rm sub}$ are translated by either $\textbf{\textit{a}}/2$ or $\textbf{\textit{b}}/2$ relative to their neighbors, as described in the text and Figure\,\ref{SrFeSe_unitcell}. Both fits employed the same background function. Hydrogen atoms attached to the nitrogen atoms were not modeled, but assumed to be directed towards the selenide ions as described in related systems \cite{BurrardNatMat12}. (f) and (g) show the local coordination environments around the Sr and N, respectively.}
\label{SrFeSe_structure}
\end{figure*}

\section{Structural characterization}

The crystal structures of the products were analyzed using synchrotron X-ray powder diffraction on beamline I11 at the Diamond Light Source, or beamline ID31 at the ESRF. Samples were sealed under an argon atmosphere within 0.5 mm diameter borosilicate glass capillaries. The diffractograms revealed a series of extremely sharp reflections that could all be indexed on a body centered tetragonal unit cell with lattice parameters ${a_{\rm sub}=2.7\,\rm\AA}$ and ${c=17.4\,\rm\AA}$ (see Fig.\,\ref{SrFeSe_structure}a). The intensities of these reflections were accounted for by the model shown in Fig.\,\ref{SrFeSe_structure}b-c in which the familiar square planar Fe nets found in all the iron selenide superconductors with Fe--Fe\,${=a_{\rm sub}\sim2.7\,\rm\AA}$  are related by the body centering translation. 

Since in this model the basal lattice parameter $a_{\rm sub}$ was equal to the Fe--Fe distance the selenide ions were inevitably modelled as disordered over two sites located above and below the centers of each square of Fe ions. Further sites were located in the interlamellar space and they were occupied by N atoms from ammonia or amide moieties (N:Fe ratio 1:2) and Sr ions (Sr:Fe ratio 0.3:2). This model was also obtained using a charge flipping algorithm implemented within {\it Topas Academic} \cite{TOPAS}, and was consistent for all the samples investigated. The interatomic distances and the coordination environments using this model were chemically realistic providing that local ordering of occupied and unoccupied Se, N and Sr sites was imposed. This required a ${a=b=\sqrt{2}a_{\rm sub}}$ basal expansion of the tetragonal unit cell to achieve a chemically realistic FeSe layer and stacking disorder along $c$ to account for the apparent smaller cell. 

Weak structured diffuse scattering was evident in the diffractograms as a result of the stacking disorder. This diffuse scattering was accounted for in a semiquantitative manner by constructing a superstructure in which layers were stacked along the $c$-axis in a way that respected the coordination environments for the intercalate species shown in Fig.\,\ref{SrFeSe_unitcell}a-b. NH$_{3}$ and NH$_{2}^-$ moieties were six-coordinate by selenide ions (a square of four in one layer and a pair in the layer above, producing an isosceles triangular prism, see Fig.\,\ref{SrFeSe_structure}g) such that N--H...Se distances were $3.7\,\rm\AA$, similar to those found in the analogous Li/NH$_3$ intercalates. Sr$^{2+}$ ions were coordinated by a triangle of selenide ions about $3.2\,\rm\AA$ apart and by N atoms from the amide or ammonia species $2.9\,\rm\AA$ apart (see Fig.\,\ref{SrFeSe_structure}f). In order to achieve these coordination environments, adjacent iron selenide layers were constrained to be translated relative to one another by $\textbf{\textit{a}}$/2 or $\textbf{\textit{b}}$/2. A model with a superstructure extending along the $c$ direction composed of 240 layers stacked randomly according to the chemical constraints captured the key features of the diffuse scattering (see Fig.\,\ref{SrFeSe_structure}d-e), although some discrepancies remain in quantitatively modeling the intensity distribution in these parts of the diffraction pattern. Attempts to explore the diffuse scattering in more detail using transmission electron microscopy were unsuccessful due to decomposition of these samples in the electron beam. This description of the structure in which well-defined layers are stacked in a disordered manner may be described as paracrystalline \cite{BurleyJACS124}. The refinements produced a Sr:Fe ratio of 0.15:1 and an N:Fe ratio of 0.5:1. The measurements conducted so far do not allow the N:H ratio to be determined with certainty so we use the formula  Sr$_{0.3}$(NH$_2$)$_{y}$(NH$_3$)$_{1-y}$Fe$_2$Se$_2$. Since the Sr/ammonia solution is reducing, the upper bound on $y$ is 0.6 yielding an Fe oxidiation state of +2. An Fe oxidation state of +1.8 which is found for other iron selenide systems would require $y=0.2$. In what follows we refer to the mixture of amide and ammonia moieties as NH$_z$ (${2.4<z<2.8}$).

The key difference between this intercalate and the previously described intercalates containing alkali or alkaline earth metal ions and ammonia molecules and/or amide ions is that the square arrangement of iron atoms in one layer is related to that in the neighboring layers by a basal plane translation of half a unit cell in either the $a$ or $b$ direction, resulting in the ${2.7\times2.7\times17.4\,\rm\AA^3}$ body centered tetragonal arrangement of Fe atoms (see Fig.\,\ref{SrFeSe_unitcell}c-f\,). This is in contrast to the arrangement of Fe atoms in all other iron-based superconductors in which the Fe atoms in two adjacent layers are related by the $c/2$ lattice vector and thus describe a primitive $2.7\times2.7\times c_{\rm sub}$ sublattice. Fig.\,\ref{SrFeSe_unitcell}c-f demonstrates how the nature of these translations, represented by dotted lines, result in a paracrystalline structure as follows. Consider starting from an identical base layer (the region below the first dotted line), in which the Fe, Se and intercalated species occupy the same crystallographic sites as those found in the Li case. To construct the next layer, each atom must be translated by either $\textbf{\textit{a}}/2$ or $\textbf{\textit{b}}/2$. In each case the Fe sites are the same but the positions of the Se, ammonia and metal species change depending on the chosen direction. As the translation occurs along a randomly chosen direction with the addition of each layer, it follows that for $n$ layers there would be $2^n$ possible configurations. Our interpretation of the structural data is that despite the random stacking, the well-separated FeSe layers themselves remain well ordered and this helps to explain our findings, described below, that robust superconductivity is preserved in these systems. 

\section{SQUID magnetometry}
Magnetic susceptibility measurements were recorded using a Quantum Design MPMS-XL superconducting quantum interference device (SQUID) magnetometer. Samples of the Sr/NH$_{z}$ intercalate (30.2 and 29.8 mg) and the Sr/ND$_{z}$ intercalate (34.0  and 30.2 mg) were filled and immobilized in gelatin capsules. Measurements were conducted in d.c.\ fields of 5 mT in the temperature range 2--150\,K after cooling in zero applied field (ZFC) and in the measuring field (FC). 

\begin{figure}[h]
\centering
\includegraphics[trim = 0cm 0cm 0cm 0.5cm, clip=true, width=7.8cm]{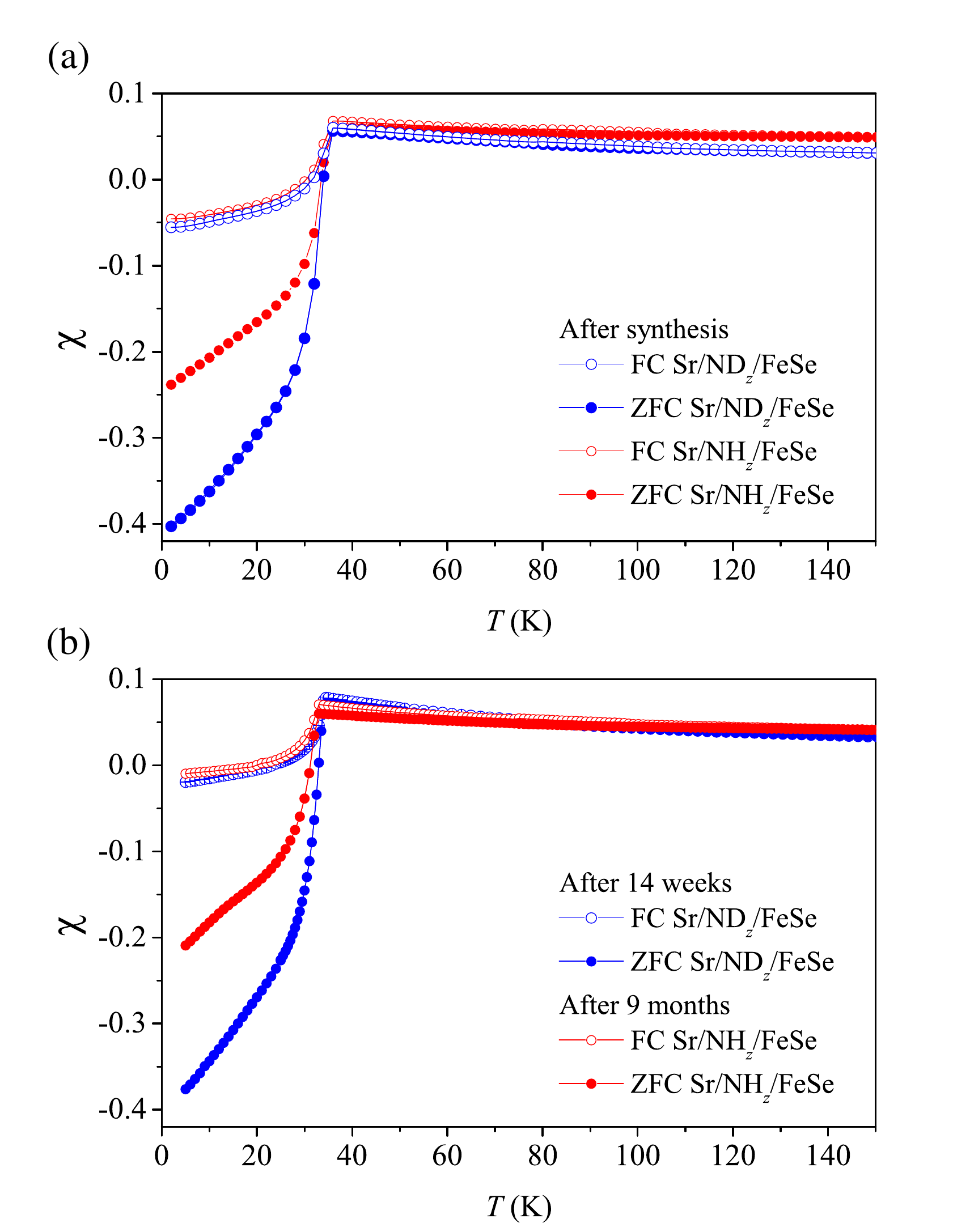}
\caption{(a) Volume susceptibility $\chi$ of Sr/ND$_{z}$ and Sr/NH$_{z}$ intercalated iron selenide after synthesis. (b) Measurements showing a slight decrease of $T_{\rm c}$ after 14 weeks and 9 months for  the Sr/ND$_{z}$ and the Sr/NH$_{z}$ intercalate, respectfully.}
\label{SrFeSe_SQUID}
\end{figure}

\begin{figure*}[t]
\centering
\includegraphics*[width=1\linewidth]{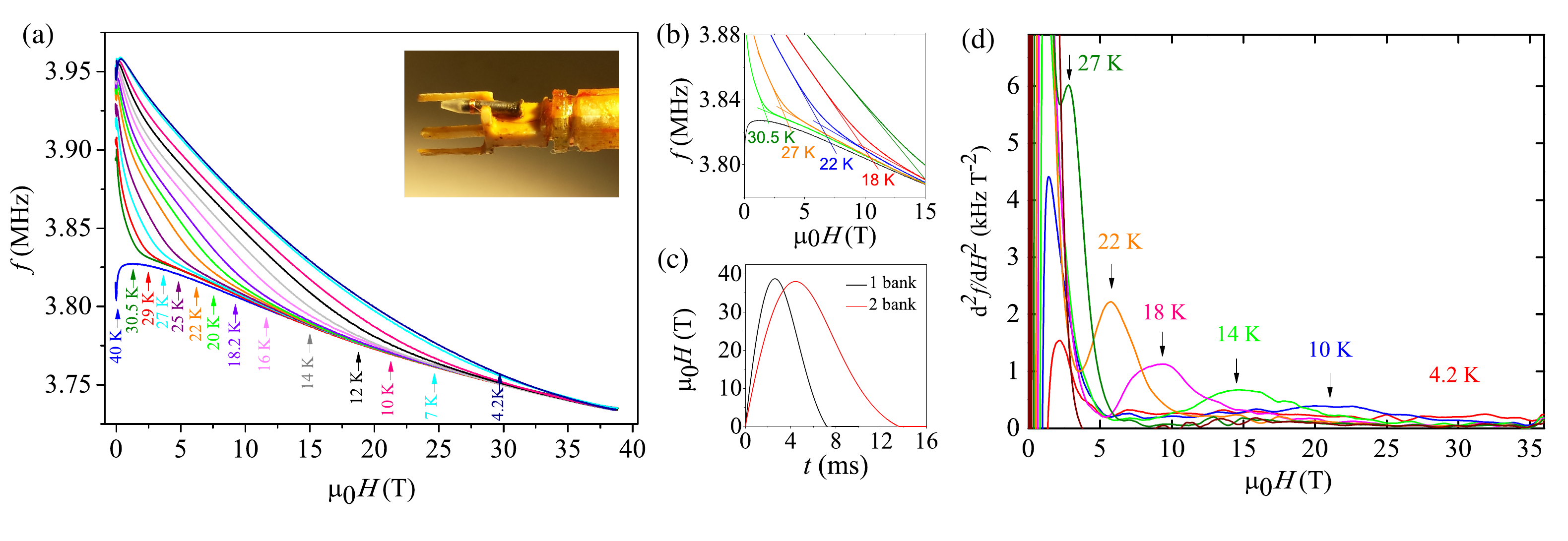} 
\caption{(a) Down-sweep portion of pulsed magnetic field data for Sr$_{0.3}$(NH$_2$)$_{y}$(NH$_3$)$_{1-y}$Fe$_2$Se$_2$ at various temperatures $T\leq40\rm~K$. Data are offset so that the frequency in the normal state matches that measured at 40~K (blue line). Inset: sample mounted on the end of the PDO probe before lowering into cryostat. (b) Close-up showing the intersection of two slopes used to determine $H_{\rm c2}$. (c) Typical 38 T field pulses demonstrate how ${{\rm d}B}/{{\rm d}t}$ varies with pulse length when using one or two banks of charging capacitors, each providing an energy of up to 0.2 MJ. (d) Second method for calculating $H_{\rm c2}$ as the peak position in ${\text{d}^2f}/{\text{d}H^2}$.}
\label{SrFeSe_powder_sweeps}
\end{figure*}

The magnetometry measurements performed directly after the synthesis (Fig.\,\ref{SrFeSe_SQUID}) indicate bulk superconductivity with a noticeable sharp drop to negative susceptibility values at ${T_{\rm c}=36(1)\,\rm K}$. There is no effect of H/D substitution on $T_{\rm c}$. Although the superconducting volume fraction is larger in the case of D, we believe this is consistent with the natural variability of volume fractions achieved in different syntheses. Even after nine months, the superconducting volume fractions and the sharp drop at $T_{\rm c}$ were found to be unchanged. A small reduction in $T_{\rm c}$ (of 2--3\,K) was detected although this was not found to be correlated with any significant change in structure. For the remainder of the paper we will present magnetometry and $\mu$SR data on the Sr/NH$_{z}$ intercalate.

\section{pulsed-field magnetometry}

Powder samples were measured at the Nicholas Kurti Magnetic Field Laboratory, Oxford using a proximity detector oscillator (PDO) dynamic susceptometer \cite{AltarawnehRevSci80, GhanRSI82}. The sample is mounted in an ampoule under argon atmosphere and placed inside a small sensor coil that is inductively coupled to the PDO circuit. This is essentially an LCR circuit with a resonant frequency {\it f} that is measured as a function of field and temperature. In metals (superconductors) {\it f} is highly dependent on changes in the skin (penetration depth) and in insulators the signal is dominated by the magnetic permeability. Thus when sweeping in field or temperature the superconducting phase transition manifests as a large change of {\it f} at $H_{\rm c2}$ or $T_{\rm c}$, which corresponds to the difference in skin and penetration depth of the two states.

Data were taken for temperatures in the range $4.2-40 \,  \text K$ in fields of up to $38\rm \,T$. Below $T_{\rm c}$ the onset of superconductivity is marked by a sudden rise in frequency and a deviation from the normal-state signal  (see Fig.\,\ref{SrFeSe_powder_sweeps}a). Additional heating can occur due to eddy currents that are generated in the mixed or normal state by the applied field. In some cases this produces hysteresis in the data and a shift in $H_{\rm c2}$, with the effect being more pronounced with a larger $\text d B / \text d t$ (a shorter pulse length for the same maximum field). We find our samples to be insensitive to these effects and observe no difference in $f$ at varying pulse lengths between {7\,ms} and {14\,ms} (see Fig.\,\ref{SrFeSe_powder_sweeps}c for field profiles of the pulses). 

\begin{figure}[h]
\centering
\includegraphics*[trim = 2cm 0.5cm 1.75cm 2.1cm, clip=true, width=7.6cm]{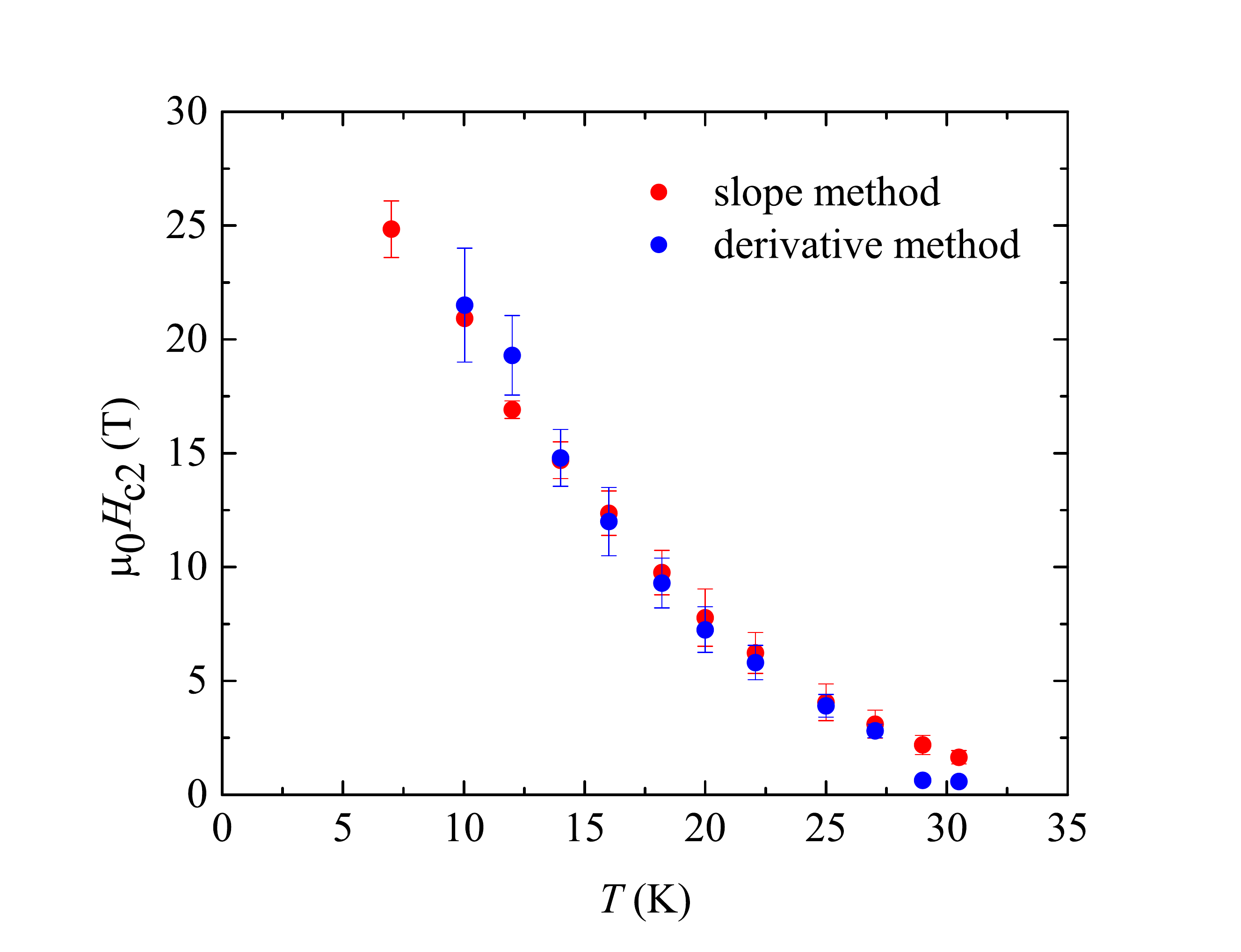} 
\caption{Phase diagram for Sr$_{0.3}$(NH$_{2})_y$(NH$_3)_{1-y}$Fe$_2$Se$_2$ using the slope and derivative methods as described in the text. The transition is too broad for a reliable estimate of $H_{\rm c2}$ at $4.2\,\rm K$.}
\label{SrFeSe_powder_phase}
\end{figure}

We note that due to the broadness of the phase transition the value of $H_{\rm c2}$ is particularly difficult to determine at low $T$. This may be due to the critical field anisotropy between the in-plane and out of plane directions increasing as the system is cooled and so for a powdered sample there is a more gradual change in frequency. Thus, we define two different methods for extracting $H_{\rm c2}$. Firstly we take $H_{\rm c2}$ as the peak in ${{\rm d}^2f}/{{\rm d}H^2}$ and its uncertainty as the half-width at half-maximum (see Fig.\,\ref{SrFeSe_powder_sweeps}d). Secondly one may perform a linear extrapolation of the curve on either side of the transition and define $H_{\rm c2}$ as the point of intersection of these two slopes. In this case, another criterion is the point at which the line extrapolated from the curve below the critical field crosses the normal-state signal. We take the uncertainty as the difference between these two points of intersection (see Fig.\,\ref{SrFeSe_powder_sweeps}b). We find that both methods give similar results with $H_{\rm c2}$ increasing steadily as the sample is cooled (see Fig.\,\ref{SrFeSe_powder_phase}). We note that the curve exhibits a concave form that has been observed in multi-band high $T_{\rm c}$ superconductors \cite{GhanPRB89}.

\section{Transverse field $\rm \mu$SR measurements }
To probe the internal field distribution in the vortex state, transverse field $\mu$SR (TF $\mu$SR) measurements were carried out using the GPS instrument at the Swiss Muon Source (PSI), Switzerland and the MuSR spectrometer at ISIS, UK. In these measurements, spin-polarized muons are implanted into the material with an external field $B_{\rm app}$ applied perpendicular to the initial muon spin direction. They will then rotate at the Larmor frequency $\omega\!=\!\gamma_{\mu} B_{\rm loc}$ where ${\sfrac{\gamma_{\mu}}{2\pi}\!=\! 135.5\,\text{ MHz T}^{-1}}$ is the muon gyromagnetic ratio and $B_{\rm loc}$ is the local field, which may include contributions from $B_{\rm app}$, magnetic ions and nuclear dipoles. This is measured directly by the time-dependent positron decay asymmetry $A(t)$ \cite{BlundellContPhy40}.

\begin{figure}[h]
\centering
\includegraphics[width=8.3cm]{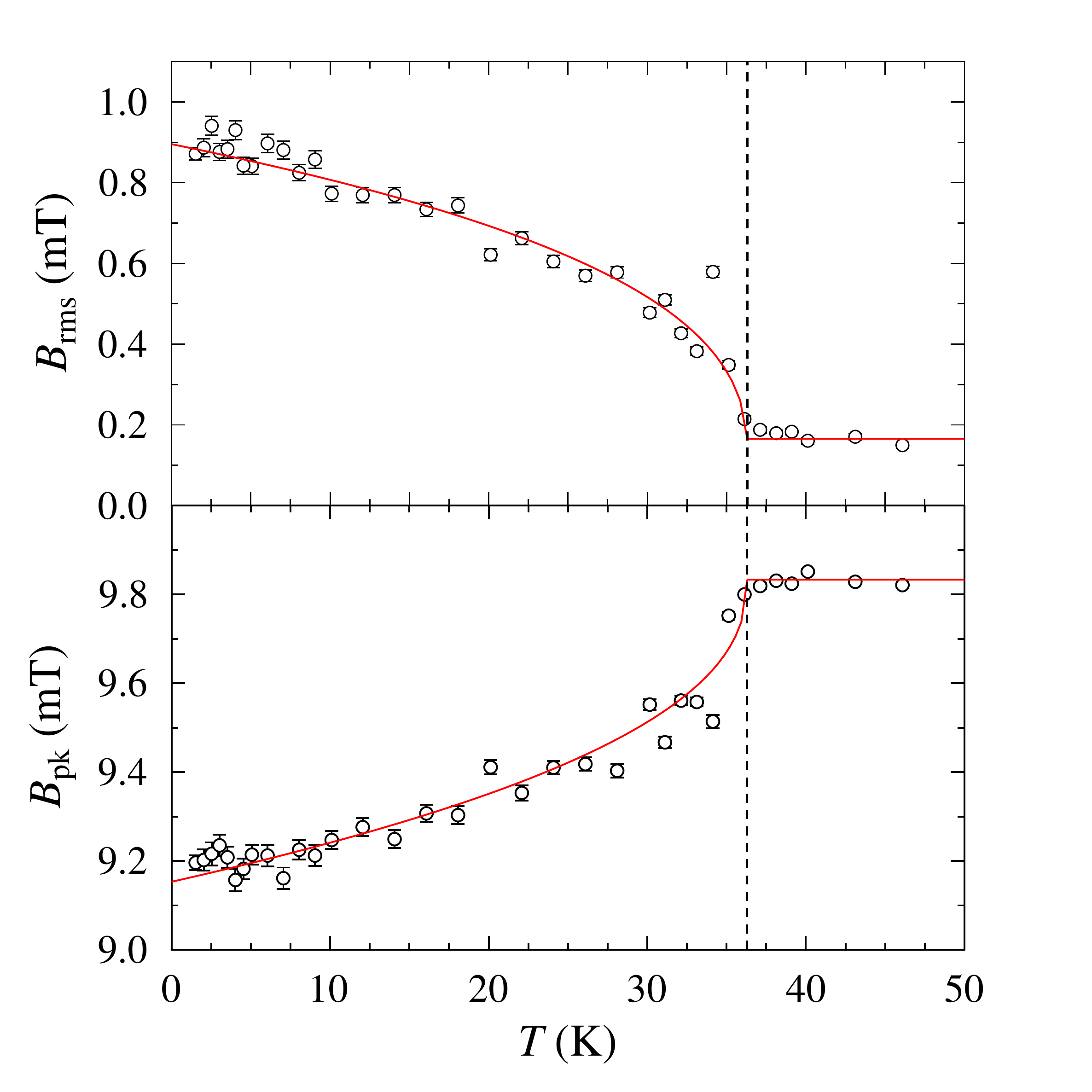}
\caption{Temperature dependences of field width $B_{\text {rms}}$ and peak field $B_{\text {pk}}$ from TF $\mu$SR measurements made on the MuSR spectrometer. Fits (red lines) are to the Eq.\,\ref{Brms(T)_fit}  and  Eq.\,\ref{Bpk(T)_fit} with the former used to extract absolute values of $B_{\text{rms}}(T=0)$ and $\lambda_{\text{ab}}(T=0)$.}
\label{TF_SrFeSe_fits}
\end{figure}

Muons implanted in a type-II superconductor will settle in particular crystallographic sites and experience a magnetic field due to the applied field as well as any internal variation resulting from the formation of a vortex lattice. However, as the vortex lattice is, in general, incommensurate with the crystal lattice the full range of the field distribution within the vortex lattice is sampled, although it is broadened slightly by the field from nuclear spins. In TF measurements, the measured asymmetry is proportional to the spin polarization as measured by individual detector banks. This experimental situation can be modeled using a fit function;
\begin{equation}\label{musr_fit}
\begin{split}
A(t) &= A_{\text {bg}}\cos(\gamma B_{\text {app}} t + \phi)\exp(-(\sigma_{\text {bg}} t)^2/2) \\
      &+ A_{\text {VL}}\cos(\gamma B_{\text {pk}} t+ \phi)\exp(-(\sigma_{\text {VL}} t)^2/2),
\end{split}
\end{equation}
where the phase $\phi$ results from the detector geometry, and  $A_{\text {VL}}$ and $A_{\text {bg}}$ are the relaxing asymmetry due to  vortex lattice and background contributions, respectively (the latter originates from muons stopping in the non-superconducting fraction of the sample or silver sample holder). This model assumes a symmetric Gaussian distribution of local fields with a peak value $B_{\text{pk}}$ ($B_{\text {app}}$) and standard deviation, or damping factor, $\sigma_{\text {VL}}$ ($\sigma_{\text {bg}}$) corresponding to the vortex lattice (background) contributions. The width of the field distribution is given by ${B_{\text {rms}}\!=\!\sigma_{\text {VL}}/\gamma_{\mu}}$ and its temperature dependence is shown in Fig.\,\ref{TF_SrFeSe_fits}. 
We observe a broadening of the field distribution (increase in $B_{\text{rms}}$) when the sample is cooled through $T_{\rm c}$ in an applied field of {10\,mT} that can be attributed to the formation of the vortex lattice with superconductivity setting in at around $36.3(2) \, \rm K$. As the material enters the superconducting state the applied field is partially screened causing a diamagnetic shift in $B_{\text{pk}}$ for $T \textless T_c$. For powdered samples of anisotropic superconductors such as these, the in-plane penetration depth $\lambda_{\text{ab}}$ is related to the field width via
 \begin{equation}\label{lambda_ab}
B_{\text{0}}=\frac{\sqrt{0.00371}\Phi_0}{(3^{\frac{1}{4}}\lambda_{\text{ab}})^2},
\end{equation}
where $\Phi_0$ is a magnetic flux quantum and $B_0$ is the vortex lattice contribution to $B_{\rm rms}$ \cite{FesenkoPhysC}. For these data it is assumed that the only other contribution to $B_{\text{rms}}$ is from nuclear dipole fields, which are temperature independent and add in quadrature. We fit our data to the phenomenological functions:
 \begin{equation}\label{Brms(T)_fit}
B_{\text{rms}}(T)=\Big\{ B_0^2 \Big[ 1-  \Big( \frac{T}{T_{\text c}} \Big)^{\alpha} \Big]^{2\beta} + B_{\text{dip}}^2\Big\}^{\frac{1}{2}}
\end{equation}
 \begin{equation}\label{Bpk(T)_fit}
B_{\text{pk}}(T)=B_{\rm app} -  B_{\rm dia}\Big[ 1-  \Big( \frac{T}{T_{\text c}} \Big)^{\alpha'} \Big]^{\beta'},
\end{equation}
where  $B_{\rm dia}$ is the maximum diamagnetic shift of the peak field, and $B_0$ and $B_{\text{dip}}$ are the widths corresponding to the vortex lattice (at ${T=0}$) and nuclear dipole contributions, respectively. For the fits shown in Fig.\,\ref{TF_SrFeSe_fits} the fitted parameters are ${\beta=0.33(3)}$ and ${\beta'=0.43(3)}$ with fixed ${\alpha\,=\,\alpha'=1}$. Using a weighted average between the extracted value of $B_0$ of these data and additional measurements on a second batch of sample we calculate a penetration depth of ${\lambda_{\text{ab}}(T=0)\!=\!292(3)\, \rm nm}$.

\section{Zero field and longitudinal field $\mu$SR measurements }

\begin{figure}[h]
\centering
\includegraphics[width=8cm]{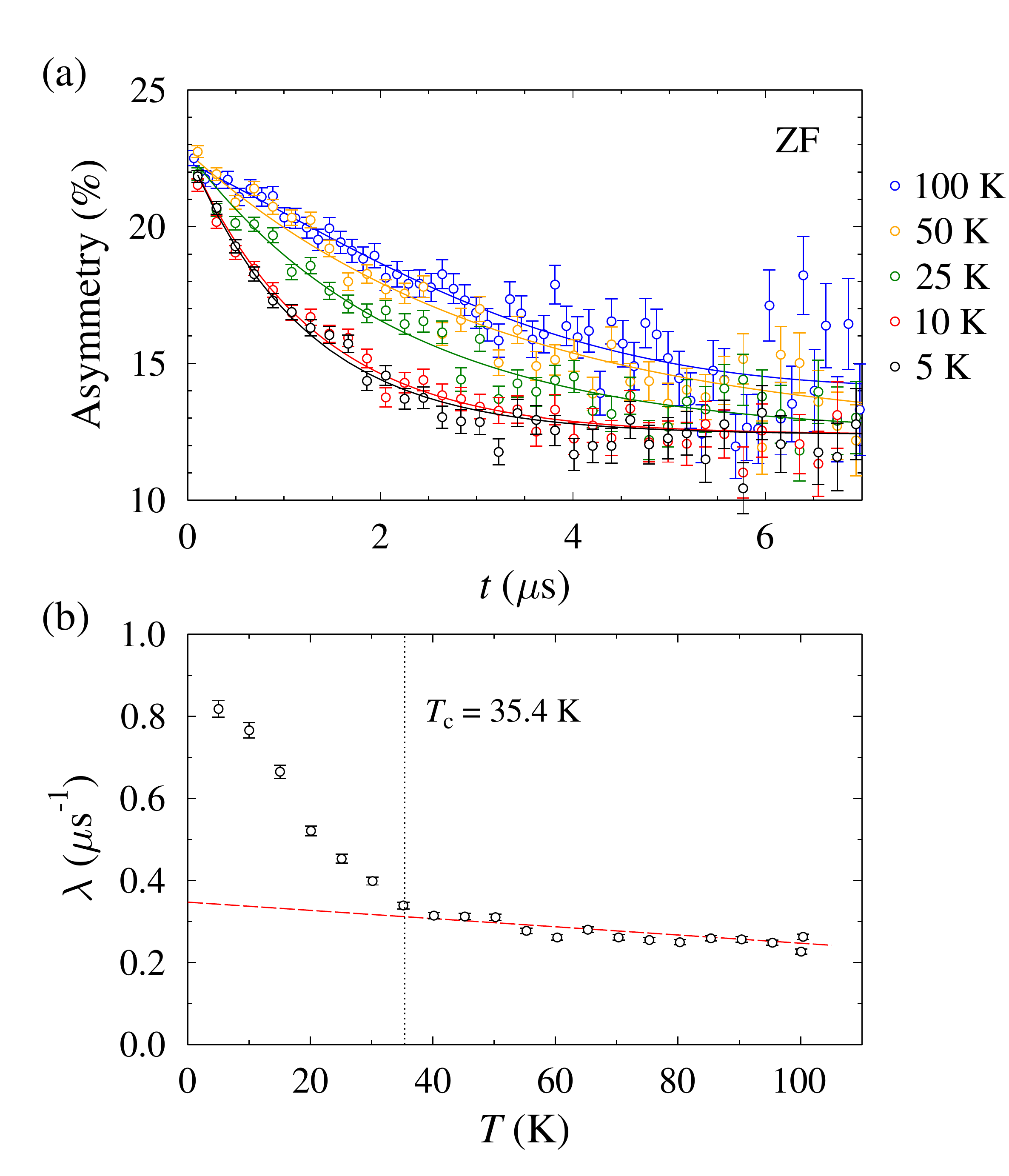}
\caption{(a)  Example asymmetry spectra from ZF $\mu$SR measurements taken on the GPS spectrometer. Fits are to Eq.\,\ref{ZF_muSR_fit}. (b) Temperature dependence of relaxation rate $\lambda$ measured in zero-field. Below $T_{\rm c}$ the data points deviate sharply from the linear trend (red line).}
\label{ZF_SrFeSe}
\end{figure}

Zero-field measurements (ZF $\mu$SR) were used to further probe the intrinsic magnetism of the system. No spontaneous oscillations were observed in the forward-backward asymmetry across the whole temperature range, nor was there any discontinuous change in amplitudes or recovery of the baseline asymmetry at low temperatures (see Fig.\,\ref{ZF_SrFeSe}a). Together these make the presence of any long-range magnetic order unlikely. Given that the spectra do not follow a Kubo-Toyabe relaxation, it is also unlikely that the relaxation is caused solely by nuclear moments. The lack of recovery of the baseline asymmetry at late times suggests dynamic fluctuations and we therefore attribute the ZF signal to relaxation caused by disordered, fluctuating electronic moments. The data were fitted to the function
\begin{equation}\label{ZF_muSR_fit}
A(t)=A_{\text{rel}}\exp(-\lambda t)+A_{\text{base}},
\end{equation}
where $A_{\text{rel}}$ is the relaxing asymmetry with relaxation rate $\lambda$ and $A_{\text{base}}$ is a non-relaxing background. $A_{\rm rel}$ can be considered as the lower bound for the superconducting volume fraction, which for this sample was around $1/3$. Exponential relaxation often describes dense spin systems that are dynamically fluctuating, where the relaxation rate $\lambda$ is proportional to the variance of the local magnetic field distribution and also to the fluctuation time, or a dilute spin systems in the presence of dynamic fluctuations. Upon cooling, the relaxation rate $\lambda$ (Fig.\,\ref{ZF_SrFeSe}b) is seen to increase slowly with decreasing temperature. However, below superconducting $T_{\rm c}$ it increases far more rapidly, suggesting that the relaxation rate is in some way coupled to the superconducting order parameter. 

\begin{figure}[h]
\centering
\includegraphics[width=8cm]{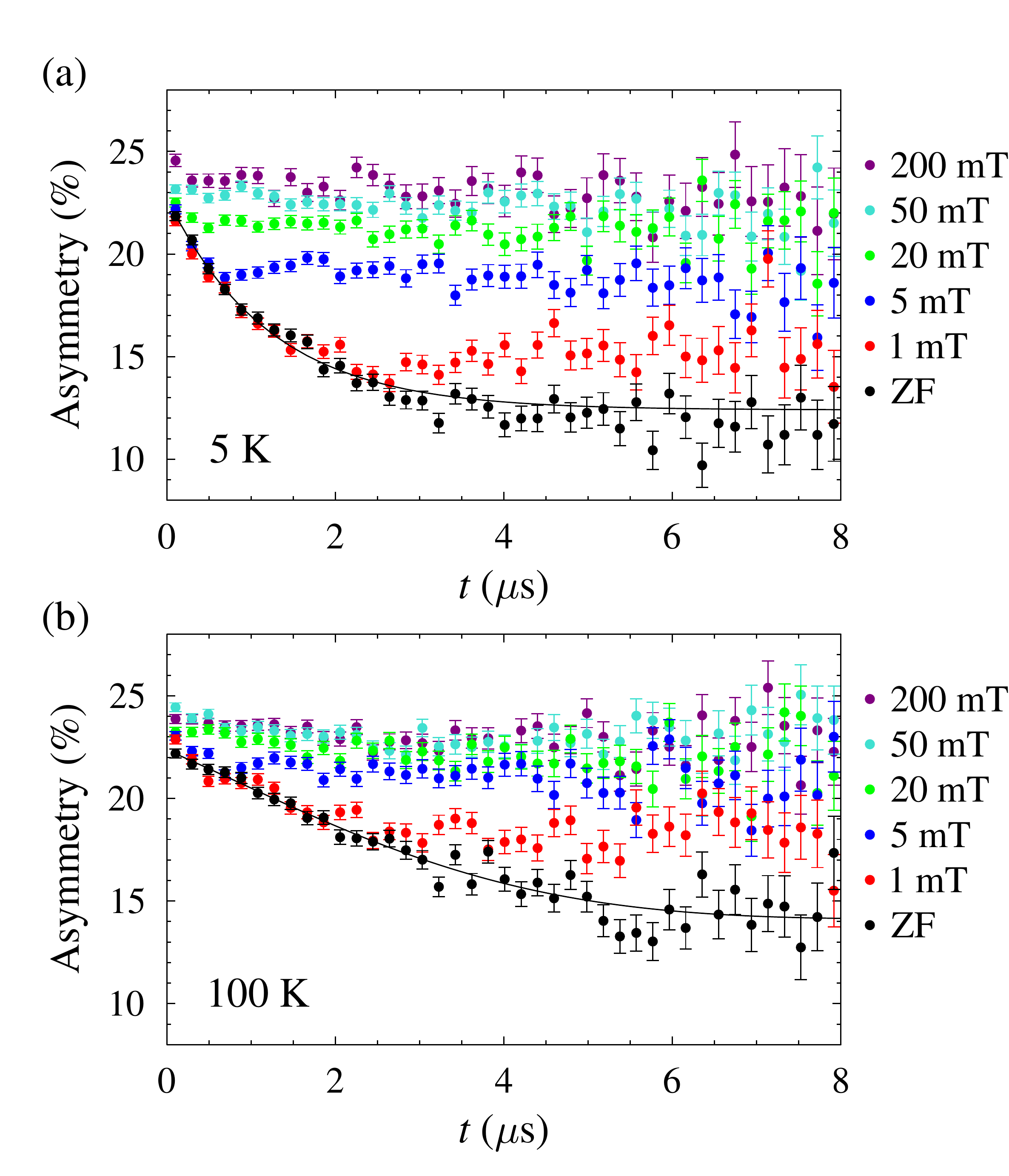}
\caption{Example LF $\mu$SR spectra at (a) 5 K and (b) 100 K. Above $T_{\rm c}$ the spins are decoupled at relatively low fields ($\approx$\,5\,\rm mT).}
\label{LF_SrFeSe}
\end{figure}

Longitudinal field measurements (LF $\mu$SR) were made at 5~K and 100~K in which a field was applied in the initial direction of the muon polarization  (see Fig.\,\ref{LF_SrFeSe}). For data measured at 5~K (Fig.\,\ref{LF_SrFeSe}a) the sample was cooled below $T_{\rm c}$ in zero applied field. At both temperatures the relaxation is decoupled at relatively low fields ($\approx 5$~mT at 100~K and $\approx 20$~mT at 5~K), suggesting that the relaxation is due to fairly dilute and/or static spins with residual dynamic fluctuations. One scenario is that there is a sizeable contribution to the ZF relaxation from static nuclear moments, with dilute, fluctuating electronic moments providing additional relaxation.  The lack of any peak in $\lambda$ suggests the absence of any freezing of the dynamics of the moments (as one would expect in a spin glass due to magnetic interactions), supporting the interpretation that these are dilute and not strongly interacting with each other. The fact that the relaxation rate is coupled to the superconducting order parameter implies that the moments are embedded in the superconductor, rather than in phase-separated pockets. At high temperatures compared to $T_{\mathrm{c}}$, the moment size and/or fluctuation rate increase slowly upon cooling. Upon cooling below $T_{\mathrm{c}}$ the muon spins likely experience a combination of slower fluctuation times, larger moments and an increased width of the field distribution. The latter could arise in a manner analogous to the increase in $B_{\mathrm{rms}}$ in the TF measurements. Specifically, if the moments are locally in a normal (rather than superconducting) region and sufficiently numerous that there is a degree of overlap of their magnetic fields in the superconducting regions, then the decreasing penetration depth upon cooling would lead muons, which decorate all of the sample, to experience a broader distribution of local fields. Taken together, the ZF measurements therefore suggest the presence of a small concentration of dilute magnetic moments dispersed in the superconducting volume, but whose presence does not seem to adversely affect the presence of superconductivity.

\section{Discussion}
Pulsed-field measurements indicate an upwards curvature in $H_{\rm c2}$ vs temperature (see Fig.\,\ref{SrFeSe_powder_phase}), which may be indicative of a multiband nature. This type of behavior has been observed in other layered two-band superconductors such as the FeAs-based systems \cite{HunteNat453, KanoJPSJ78}, and the parent compound FeSe using resistivity measurements \cite{AmigoJPCS568}. Fits of the Werthamer-Helfand-Hohenberg (WHH) model \cite{WerthamerPR147} for a one-band superconductor in the dirty limit did not converge. Instead we use a model developed by Gurevich that is based on a weakly-coupled two-band BCS superconductor \cite{GurevichPhysC456, GurevichPRB67}. This includes scattering from non-magnetic impurities, orbital pair breaking, strong electron-phonon coupling and spin paramagnetism (note that unlike the WHH model, it does not account for spin-orbit effects). The model parameters are the band diffusivities $D_1$ and $D_2$, and the intraband ($\lambda_{11}$ and $\lambda_{22}$) and interband ($\lambda_{12}$ and $\lambda_{21}$) coupling constants. Assuming that the non-magnetic impurity scattering does not affect $T_{\rm c}$ the upper critical field is given in reduced natural units ${h = {H_{\rm c2}D_1}/{2 \phi_0 T}}$ and $t=T/T_{\rm c}$ by
%
%
\begin{equation}\label{Gurevich_model}
\begin{split}
a_0 [ \ln{t} &+ U(h)] [\ln{t} + U(\eta h)] \\
&+ a_2[ \ln{t} + U(\eta h)] + a_1[ \ln{t} + U(h) ] = 0,
\end{split}
\end{equation}
where
\begin{equation}\label{Gurevich_model2}
U(x) = \psi \bigg( \frac{1}{2} + x + i \frac{\mu_{\rm B}H}{2 \pi T} \bigg) - \psi(x), 
\end{equation}
in which $\psi(x)$ is the digamma function, ${\eta = D_2/D_1}$ is the ratio of band diffusivities, and the constants $a_0,$ $a_1$ and $a_2$ are functions of the coupling constants.

\begin{figure}[h]
\centering
\includegraphics*[trim = 0cm 0.5cm 0cm 2cm, clip=true, width=8.5cm]{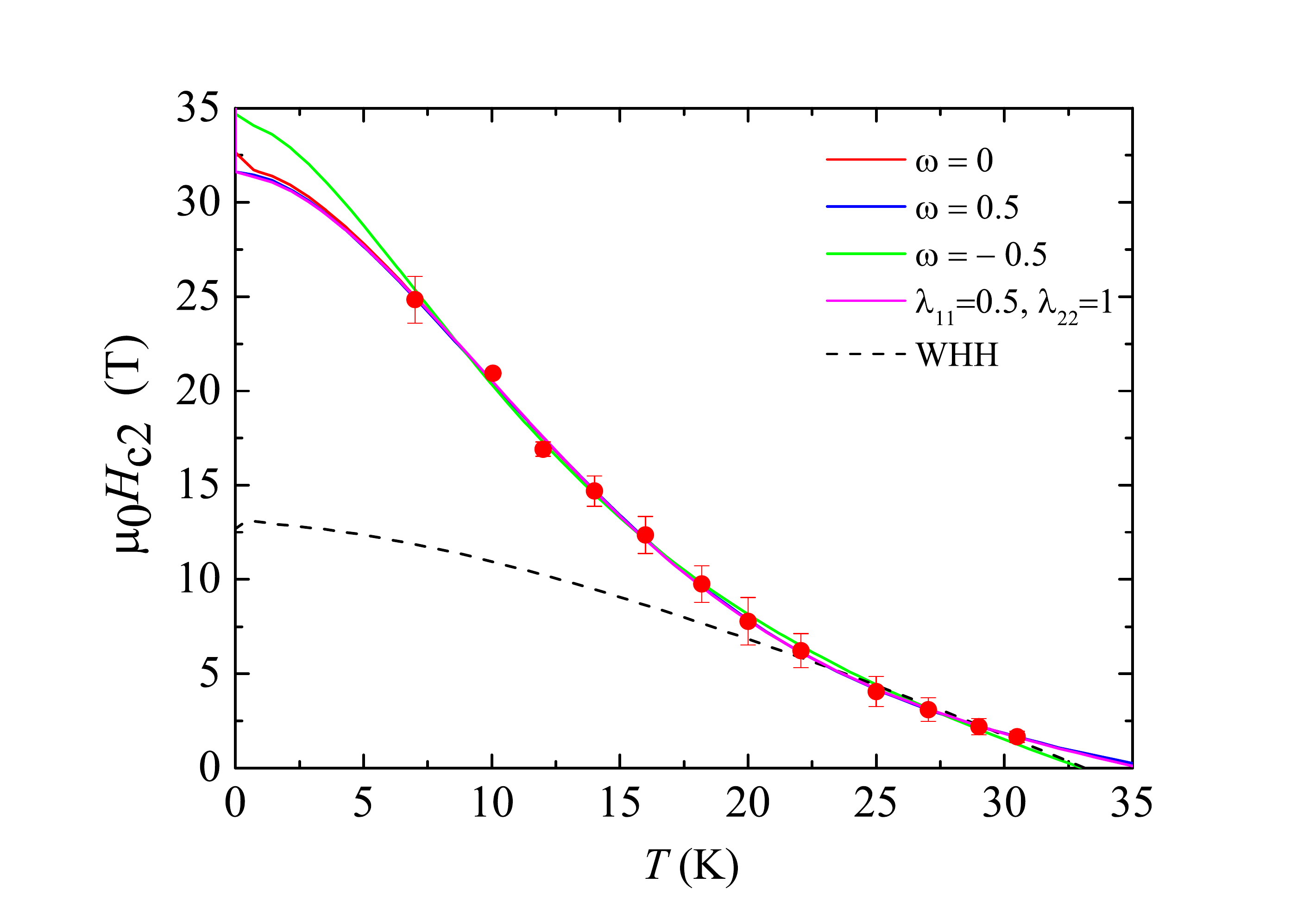} 
\caption{Fits of $H_{\rm c2}$ using the two-band Gurevich model (Eq.\,\ref{Gurevich_model}) with either a fixed $\omega$ or fixed intraband coupling constants $\lambda_{11}$ and $\lambda_{22}$ ($H_{\rm c2}$ determined by the slope method). Dotted line shows a fit to the single-band WHH model.}
\label{Gurevich_fits}
\end{figure} 

In any case, the model is difficult to fit due to overparameterization; at high temperatures, $H_{\rm c2}$ is strongly dependent on $D_2$/$D_1$ and weakly dependent on the coupling constants, which contribute significantly only in the mK temperature range. Consequently, fits to the data converge only when one set of coupling constants (either interband or intraband) are fixed. Fits are shown in Fig.\,\ref{Gurevich_fits} for various values of $\lambda_{11}, \lambda_{22}$ and ${w=\lambda_{11}\lambda_{22}-\lambda_{12}\lambda_{21}}$. The Gurevich model was successful in modeling the upwards curvature of $H_{\rm c2}$ which stems from the difference in diffusivities of the two bands i.e. ${\sfrac{D_2}{D_1}<1}$ \cite{GurevichPRB67}. However, we found that it was possible to fit for both a strong (when ${w<0}$) and weak (${w>0}$) interband coupling and fitted parameters varied considerably depending on the values of the chosen coupling constants and the method for finding $H_{\rm c2}$. This is particularly noticeable when extrapolating $H_{\rm c2}(T)$ to absolute zero. When using $H_{\rm c2}$ values extracted by the slope method as described earlier, we find ${\mu_0H_{\rm c2}(0)\!\approx\!33(2)\rm\,T}$ to be fairly consistent across various different scenarios (see Fig.\,\ref{Gurevich_fits}). The derivative method for calculating $H_{\rm c2}$ could not be extended to the lowest temperature datum and hence produced a poorly constrained estimate, thus this method was deemed less reliable.

\begin{figure}[h]
\centering
\includegraphics*[trim = 0cm 0cm 0cm 0cm, clip=true, width=7.5cm]{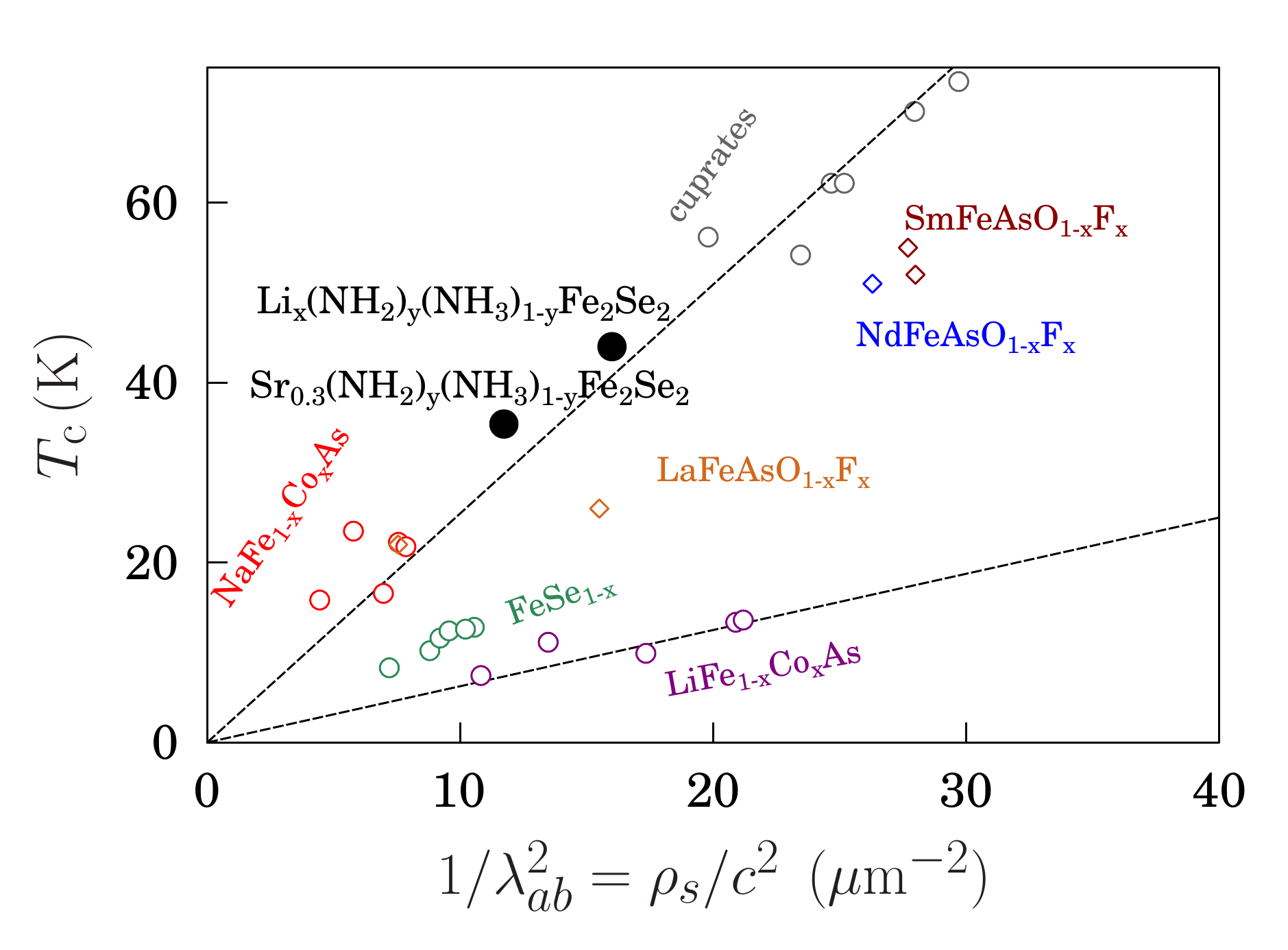} 
\caption{The Uemura plot of $T_{\rm c}$ against superfluid stiffness $\rho_{\rm s}=\sfrac{c^2}{\lambda_{\rm ab}^2}$ shows that the intercalates $A_{\text x}$(NH$_2$)$_{y}$(NH$_3$)$_{1-y}$Fe$_2$Se$_2$ (${A=\rm Sr, Li}$) fall close to the main scaling line.}
\label{Uemura_plot}
\end{figure} 

The Uemura relation \cite{UemuraPRL62} is a scaling relation between $T_{\rm c}$ and the superfluid stiffness $\rho_{\rm s}$ which seems to hold well for many exotic superconductors \cite{UemuraPRL66}. It is known to break down for overdoped cuprates \cite{NiedermayerPRL71} and may be an over simplification \cite{TallonPRB68}, and other scaling relation behaviours have been explored \cite{HomesNat430,PrattPRL94,TallonPRB73,TaylorPRB76}.  Nevertheless, the Uemura plot of $T_{\rm c}$ against $\rho_{\rm s}$ provides a convenient means of exploring the energy scale to break up pairs as a function of the strength of the order parameter, and our TF $\mu$SR data allow us to extract an estimate of $\rho_{\rm s}=c^2/{\lambda^2_{ab}}$.  As shown in Fig.~10, we find that the Sr intercalated compound is close to the main scaling line on the Uemura plot, as is the Li intercalated material \cite{BurrardNatMat12}, and this behavior correlates with underdoped cuprates and many other iron-based superconductors.  Note that there is also another, lower line in this plot, which is common to electron-doped cuprates \cite{HomesPRB56,ShengelayaPRL94} and LiFe$_{1-x}$Co$_x$As \cite{PrattPRB79,PitcherJACS132}. For those materials, it is found that although the superconducting state  is reasonably robust (the superfluid is stiff) the strength of the pairing is significantly suppressed, but these considerations do not seem to apply to our intercalated compounds.

\section{Conclusion}

We have probed the superconducting properties of a recently discovered superconductor Sr$_{0.3}$(NH$_{2})_{y}$(NH$_3)_{1-y}$Fe$_2$Se$_2$ using pulsed-field magnetometry and $\mu$SR techniques. The upper critical field was shown to increase upon cooling and exhibited a concave form that is reminiscent of other multigap high-temperature superconductors, which when extrapolated to absolute zero gives a maximum upper critical field of ${\mu_0H_{\rm c2}(0)\!\approx\!33(2)\rm\,T}$. TF $\mu$SR measurements show a clear diamagnetic shift and a broadening of the field width that is highly reproducible between different batches of sample, with the onset of superconductivity at $36.3(2) \, \rm K$. ZF $\mu$SR did not reveal any long-range magnetic order but dilute electronic moments with some residual dynamics and whose behavior is coupled to the superconducting order parameter. We find that intercalation of Sr atoms together with amide and ammonia introduces intrinsic stacking disorder that results in a paracrystalline state. The system nevertheless retains complete structural order of the Fe sublattice, and thus these results demonstrate that robust superconductivity does not rely on perfect structural coherence along the $c$-axis. 

\begin{acknowledgments}
This work was supported by the UK EPSRC  via grant number EP/I017844. We thank the EPSRC for studentship support for FRF and fellowship support for TL. We thank the STFC ISIS Facility and Swiss Muon Source, Paul Scherrer Institut for the provision of beam time and are grateful for the assistance of A. Amato, P. J. Baker and instrument scientists on beamline I11 at Diamond.  SJC thanks both EPSRC and Diamond for studentship support. SJS acknowledges the support of a DFG Fellowship (SE2324/1-1). We are grateful to Prof J. Hadermann (EMAT, Antwerp) for attempting transmission electron microscopy measurements on these samples.
\end{acknowledgments}

\bibliography{SrFeSe_paper}

\end{document}